# Investigating meV-scale Equilibrium Atomic Dynamics with X-Rays: focus on disordered materials

Alfred Q.R. BARON

baron@spring8.or.jp

*Materials Dynamics Laboratory, RIKEN SPring-8 Center 1-1-1 Kouto, Sayo, Hyogo, 679-5148 Japan*

This paper reviews non-resonant meV-resolved inelastic x-ray scattering (IXS) as a complementary method to inelastic neutron scattering (INS). Two aspects of IXS that are notable in this context are (1) that IXS allows straightforward measurements of phonons in small (sub-mm, and even ~0.01 mm) samples with ~1 meV resolution and excellent Q resolution and (2) that IXS avoids the kinematic constraints of INS. The first allows both new geometries (thin films, diamond anvil cells) and easy access to new materials, while the second allows high-quality data on disordered materials – e.g. scans with sub-meV resolution to arbitrarily high energy transfer are possible, even at Q = 1nm$^{-1}$. The review briefly discusses the spectrometers and compares the practical forms for the dynamic structure factor, $S(Q,\omega)$, for crystals, glasses and liquids. This is followed by a longer review of work on liquids and then shorter discussion of work on glasses, crystals, and in specific geometries. This paper complements the longer introduction/review of IXS at http://arxiv.org/abs/1504.01098 .

## 1. Introduction

This paper reviews measurements of meV-scale (THz frequency or ps time scale) equilibrium atomic dynamics using x-rays, emphasizing how meV-resolved inelastic x-ray scattering (IXS) complements inelastic neutron scattering (INS). It is hoped this will be useful to scientists having a background in neutron scattering, as is the expected readership of (the original version of) this article appearing in a publication of the Japanese Society for Neutron Science. We emphasize that the focus is specifically on non-resonant measurements where the x-ray energy is chosen to achieve the best resolution. This means that the dominant interaction occurs via Thomson scattering, and the method is not sensitive to spin excitations.

The meV and (nm to) Å scales discussed here are exactly the scales of atomic motion relevant to understanding the equilibrium dynamics of most condensed matter systems. Thus there is a general need to probe excitations on these scales as a basic spectroscopy for investigating material properties including *elasticity, superconductivity, thermal transport, thermoelectricity, ferro-electricity, multi-ferroicity, metal-insulator transitions, viscosity, fragility, relaxations and dispersion in disordered materials, behavior near critical points*, *etc*. Further, atomic dynamics have a role in many phase transitions, both as related to the previous list of properties, and as may be relevant to structural changes and the appearance of charge density waves. There is also an increasing interest in the detailed interaction of phonons with other systems, including electron-phonon coupling (e.g. Kohn anomalies, Peierls instabilities, and generalizations there-of), magnon-phonon coupling, and interaction with other quasi-particles. This takes on added importance in an environment where researchers would like to tailor the response of functional materials by balancing interactions between multiple systems (phononic, electronic, magnetic). Other recent relevant topics include the topology of phonon dispersion bands (e.g., Dirac points) and phonons carrying angular momentum. All of these drive an interest in – even a necessity of - probing equilibrium atomic dynamics at meV scales.

This article will discuss some of the complementary aspects of INS and IXS, emphasizing the opportunities with IXS, then briefly discuss the operating points of IXS spectrometers. Then we will review IXS work covering disordered materials, crystalline materials and some specific geometries – including measurements of samples in diamond anvil cells (DACs) and of thin films. The paper focusses most heavily on liquid dynamics as the other topics are at least partly covered in recently updated review articles. Finally, we close by considering some perspectives for future work. We will focus on work done in Japan to help limit the scope of the paper and to point out the local opportunities.

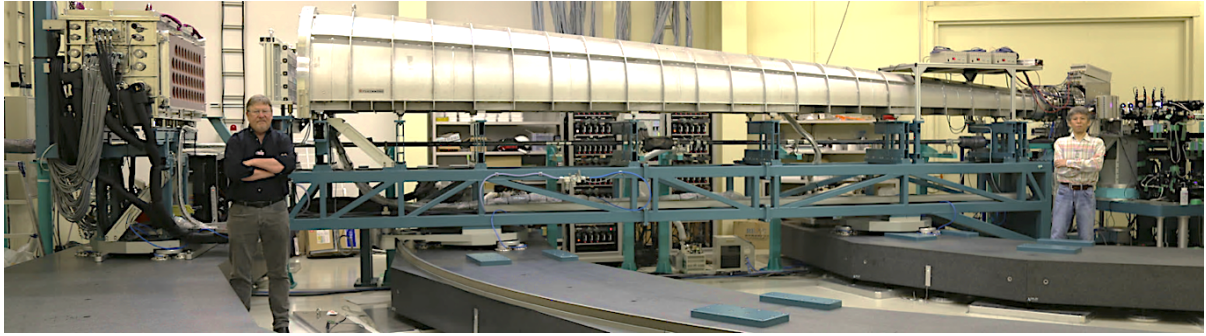

**Figure 1.** Photo of the 10m two-theta arm of the meV IXS spectrometer at BL43LXU of SPring-8. The x-ray bean is incident from the right and the sample location is at the far right of the photo, while the backscattering analyzer array is in the vacuum chamber at the far left. The arm moves on air pads on the granite base. The arm at BL35XU is very similar but uses cylindrical vacuum chambers.

The presentation here attempts to be clear, direct, broad, rigorous and to keep the readers' interest. It is, of course, difficult to succeed in all of those endeavors simultaneously, and one knows all too well that what one person chooses to skip or cover quickly may be a major part of another person's career: one hopes readers will view this work positively. For those desiring additional information some other reviews or texts that this author considers useful include [1,2] for IXS instrumentation and for IXS measurements of crystalline materials, [3] and [4] for discussion of liquids and liquid dynamics and [5] for a general background on scattering concepts. However, readers are strongly encouraged to do their own literature searches on the specific topics that may interest them, or, if relevant, to contact the author directly.

The present paper will focus on *equilibrium* dynamics as captured in measurements of the dynamic structure factor, $S(Q,\omega)$. However, the recent availability of x-ray free electron lasers, such as LCLS at Stanford or SACLA at SPring-8, which provide intense x-ray pulses, has made it possible to investigate phonons via pump-probe experiments [6] : a short laser pulse excites a system and then x-ray pulses from the FEL are used to stroboscopically probe changes in the diffuse scattering with ~fs resolution, as may be then related to phonon properties. This is not an equilibrium measurement, as it necessarily starts with a pulse that perturbs phonon populations, creating a "coherent phonon" at t=0, but can yield similar information, and in some cases more information Especially impressive was an experiment that selectively excited one mode and then observed how that mode decayed while other modes increased [7] – ie: it was possible to directly observe the anharmonic decay path of the coherent phonon.

## 2. Comparing X-Rays and Neutrons

IXS offers some unique / complementary opportunities beyond what is available using INS. This is discussed in detail in [1,2] and here we paraphrase and update part of that discussion. The two most important advantages of IXS are (1) access to small samples and (2) removal of kinematic constraints which hamper INS investigations of disordered materials. The first is the direct result of the high brilliance of modern x-ray sources: the small source size and small divergence of synchrotron sources makes it easy to focus the x-ray beam to a transverse size of ~0.1 mm or even, with some losses, to ~0.005 mm. This allows comparable size samples to be investigated, including samples in extreme (e.g. near earth's core) conditions in diamond anvil cells (DACs), multilayer or thin film structures at grazing incidence, and small crystals of new materials. The last is especially important as one can investigate phonons without having to invest huge effort to grow (or assemble) large samples, making it relatively easy to investigate the phonons of a newly discovered material. The removal of kinematic constraints results from the fact that the >~10,000 eV energy of the x-rays used is many orders of magnitude larger than <0.1 eV phonon energy transfers, decoupling energy and momentum transfer in IXS. Scans at small momentum transfers can then be made to arbitrarily high energy transfers with reasonable energy resolution. This provides access to a region of energy-momentum space that is important for understanding disordered materials (liquids and glasses) and is difficult to measure using INS.

Neutrons are advantageous when very high-energy resolution is needed, as INS spectrometers can provide <100 μeV resolution at least for smaller energy transfers. In addition, whereas the energy resolution for most meV x-ray spectrometers is approximately Lorentzian, so has long tails, the resolution with neutrons often has shorter tails. This can make neutrons advantageous for observing weak modes near to stronger ones, or for measuring modes in the presence of strong elastic backgrounds. Neutrons are extremely interesting when large single crystals of heavier materials are available, whereas x-rays are limited by the short penetration length into the sample due to the high photoelectric absorption. Also, modern time-of-flight neutron spectrometers allow collection of a huge swath of momentum space at one time, so, assuming modest (~0.1 cc) size samples are available, may offer advantages for survey measurements.

There are a variety of other complementary aspects of IXS and INS, as discussed in more detail in [1,2]. We finish this section with three additional comments. (1) IXS spectra are often relatively clean/simple compared to INS spectra as IXS has no background from either incoherent or multiple scattering, or from magnetic excitations. (2) x-ray penetration lengths are limited by strong photo-electric absorption and are usually relatively short (10 to 100 microns) compared to those for neutrons so, in most x-ray experiments, one only sees the near-surface region of a sample. (3) While it is possible to use meV-resolved IXS also to investigate *electronic* excitations if they appear at high (eV-scale) energies (see [1,2] ), in response to several recent inquiries, we note it is ***practically* not possible *investigate electronic excitations that occur at energies similar to 1-phonon scattering***: the electronic excitation cross section is just too small – generally <<1% of the phonon cross section.

## 3. meV IXS Spectrometers

There are two meV IXS spectrometers at SPring-8 in Japan, BL35XU [8] and BL43LXU [9] while outside Japan, there are two spectrometers in the United States, one at APS and one at NSLS-II, and one spectrometer at the ESRF in France. In Japan, BL35XU was built first and then a stronger facility was built by RIKEN at BL43LXU. BL43LXU also can provide better (sub-meV) resolution (see table 1) [10]. A photo of the two-theta arm at BL43LXU is shown in figure 1 and presents the scale of the spectrometers - as is reminiscent of neutron instruments. The beamlines in Japan are now the leading meV-IXS spectrometers world-wide [1,2], with BL35XU providing more flux and analyzers than any other facility except BL43LXU. The workhorse setup at the beamlines provides resolution between 1.2 and 1.5 meV FWHM, and a higher flux setup with 3 meV resolution is also available (see table 1).

Table 1: IXS beam size and energy resolution

| Beamline | Focused Beam Size | Resolution (FWHM) |
|---|---|---|
| BL35XU | <80 μm or ~20 μm | 3 meV @ 17.8 keV 1.4 meV @ 21.7 keV |
| BL43LXU | ~50 μm or ~5 μm | 3 meV @ 17.8 keV 1.2 meV @ 23.7 keV 0.8 meV @ 25.7 keV |

Both spectrometers rely on backscattering monochromators and analyzers using the Si (nnn) series of reflections operating at Bragg angles within ~0.3 mrad of exact backscattering. The operation very close to exact backscattering is what drives the large size of the two-theta arms: a long path length is needed to separate the x-rays scattered from the sample and those after the analyzer. One of the notable features of the SPring-8 spectrometers is their 2-dimensional analyzer arrays (facilities outside Japan have only a (1D) line of analyzers). These 2D arrays allow direct access to a large area of momentum space and provide simultaneous access, in a single measurement, to dispersion in both longitudinal and transverse dispersion of modes in crystalline materials (see also more detailed discussion in [1,2] ). We also emphasize that preserving the high resolution requires precise (sub-mK) temperature control: <0.2 mK rms control is routinely achieved over ~week time scale [10]. Additional information about the beamlines and spectrometers may be found in [1,2] and at http://bl35www.spring8.or.jp/ and http://beamline.harima.riken.jp/en/bl_info/bl43lxu_info.html with, a variety of practical advice on experiment design at https://beamline.harima.riken.jp/bl43lxu/

## 4. The Dynamic Structure Factor

We briefly discuss some formal aspects of the cross section as this brings out interesting contrasts between the dynamics of different material types - crystals, glasses, and liquids - and because it helps define the concepts more clearly. The cross section for scattering of a photon from an initial state with wave vector $\boldsymbol{k_1}$ and polarization $\boldsymbol{\varepsilon_1}$ to a final state $\boldsymbol{k_2}\boldsymbol{\varepsilon_2}$ is usually written as

$$\left.\frac{\partial^2 \sigma}{\partial E \partial \Omega}\right|_{\mathbf{k_1}\varepsilon_1 \to \mathbf{k_2}\varepsilon_2} = \frac{k_2}{k_1} r_e^2 \left|\varepsilon_2^* \cdot \varepsilon_1\right|^2 S(Q,\omega) \quad (1)$$

where $r_e$ is the classical radius of the electron (to be replaced by a scattering length for neutron scattering). For x-ray scattering, which is fully coherent, the dynamical structure factor, $S(Q,\omega)$, is the space-time Fourier transform of the van-Hove pair correlation function, $G(r,t)$ [11] . The formulation of eqn. (1) removes, factors out, a scale factor due to the probe-sample interaction, for x-rays, the Thomson scattering, and allows one, somewhat, to focus on material properties independent of probe characteristics. $S(Q,\omega)$ can also be expressed as the imaginary part of a linear response function (see, e.g., [12] ).

The practical expressions for $S(Q,\omega)$ for crystalline, glassy and liquid states of matter are interesting and useful to contrast the dynamics of the different systems. For crystals, to a good approximation in many cases, the dynamic structure factor may be taken to be that for 1-phonon scattering from a harmonic solid as a sum over the 3N modes (where N is the number of atoms in a primitive cell) (see [5] and [1,2] ) A crystal, of course, is not isotropic, so we need to consider the full vector nature of the momentum transfer relative to the crystal orientation. Taking $F_{1P}(\mathbf{Q},j)$ as the one-phonon structure factor of mode j at frequency $\omega_{qj}$ one has:

$$S(\mathbf{Q},\omega)_{1P} \propto \sum_j |F_{1P}(\mathbf{Q},j)|^2 \times \qquad (2)$$
$$(\langle n+1 \rangle \delta(\omega - \omega_{\mathbf{q}j}) + \langle n \rangle \delta(\omega + \omega_{\mathbf{q}j}))$$

$$|F_{1P}(\mathbf{Q},j)|^2 \equiv \frac{1}{\omega_{\mathbf{q}j}} \times$$
$$\left| \sum_d \frac{f_d(Q)}{\sqrt{M_d}} e^{-W_{Qd}} \mathbf{Q} \cdot \mathbf{e}_{\mathbf{q}jd} e^{i\mathbf{Q} \cdot r_d} \right|^2 \qquad (3)$$

The 1-phonon structure factor includes a sum over all atoms in the unit cell weighted by the atomic form factors, $f_d(Q)$, masses, $M_d$, Deby-Waller factors, $e^{-W_{Qd}}$ and the phased of the motion (phonon eigenvector, $e_{qjd}$) projected onto the total momentum transfer. The contribution of each mode is straightforward to calculate in principle, but the Q dependence can be non-obvious for optical modes due to the phased sum in eqn. (3). We note that, practically, many measurements will also show some elastic intensity due to sample imperfections – e.g. poor surface quality or other defects.

As Q and temperature are increased, the contribution from multi-phonon scattering (simultaneous annihilation/creation of more than one phonon in a single scattering event – $S(\mathbf{Q},\omega)_{2P}$, $S(\mathbf{Q},\omega)_{3P}$ etc.) will generally increase and this can to lead to additional contributions to the spectra [13] , especially near 0 energy transfer [1,2] . We also note that while most phonon linewidths are negligible on the meV scale, hence the use of the delta functions in eqn. (2), sometimes the linewidths can be non-negligible due, e.g., to strong anharmonicity or strong electron-phonon coupling. In that case the term in parentheses in eqn. (2) can sometimes be replaced by a lowest-order approximation, a damped harmonic oscillator (DHO) [14] (see also discussion in [1,2] ) as we also use below – however, the real line-shapes in crystals can be much more complicated than simple DHOs [15,16] .

For disordered materials, one must generally explicitly consider broadened phonon lines as their line widths are not negligible on a meV scale. However, the response measured from a macroscopic size sample is isotropic, so one only needs to consider the magnitude of the momentum transfer and we drop the vector (boldface) $Q$. The cross sections (at modest Q, say Q < 30 nm$^{-1}$) can usually be taken as a sum over damped-harmonic oscillator (DHO) terms for the propagating modes:

$$S(Q,\omega)_{Glass} \propto DB(\hbar\omega/k_B T) \times$$
$$\begin{pmatrix} f\delta(\omega) + \\ \frac{1-f}{\pi} \sum_j \frac{2A_j \Gamma_j \Omega_j^2}{(\omega^2 - \Omega_j^2)^2 + 4\omega^2 \Gamma_j^2} \end{pmatrix} \qquad (4)$$

$$S(Q,\omega)_{Liq} \propto DB(\hbar\omega/k_B T) \times$$
$$\begin{pmatrix} \frac{I_0}{\pi} \frac{\gamma}{\omega^2 + \gamma^2} + \\ \frac{1-I_0}{\pi} \frac{2\Gamma_B \Omega_B^2}{(\omega^2 - \Omega_B^2)^2 + 4\omega^2 \Gamma_B^2} \\ + \frac{I_0 \gamma}{\pi} \frac{\Omega_B^2 - \omega^2}{(\omega^2 - \Omega_B^2)^2 + 4\omega^2 \Gamma_B^2} \end{pmatrix} \qquad (5)$$

Here $\Omega$ and $\Gamma$ are the (dressed) energy and (~half) width of the mode having an intensity given by $(1-f)A_j$, eqn. (4), or $1-I_0$, eqn (5). In glasses, relaxations generally occur on very long (~100 s) time scales so at ~meV resolution they are described by a delta function with a relative weight given by the non-ergodicity parameter, $f$. For liquids, relaxations are often on ps or sub-ps time scales, as is not negligible at meV resolution, so the delta function must be replaced by a Lorentzian having a finite width, $\gamma$. For the liquid form we focus specifically on the acoustic or Brillouin mode as that usually dominates the inelastic response at low Q. We also note that liquids have an interaction between the relaxation (quasi-elastic mode) and the propagating mode, as given by the third term (see discussion in [17] ). For both liquids and glasses, one includes a detailed balance factor which is usually taken in the 1-phonon limit, $DB(x) = x(1-e^{-x})^{-1}$. The form given for the liquids is functionally equivalent to the usual Rayleigh-Brillouin triplet of hydrodynamics [3]. However, this should be approached carefully: at small enough Q (e.g. Q<<0.1 nm$^{-1}$) parameters can behave as expected from hydrodynamics, with the central peak due to thermal relaxation and the linewiths scaling as $Q^2$, but in the region of momentum space probed by IXS (typically $Q>1$ nm$^{-1}$) the quasi-elastic line is usually not the thermal relaxation of hydrodynamics (that, e.g., obeys the Landau-Placzek relation) but a structural relaxation that can also impact the dispersion and the width of the acoustic mode [18] – ie: the lineshape can be well described by eqn. (5) but the parameters do not have the relation to macroscopic quantities given by hydrodynamics. More generally, there can be multiple relaxations so eqn. (5) should be considered as a simplest case.

## 5. Dynamics of Disordered Materials

The experimental investigation of disordered materials – liquids and glasses - made a big step forward with the development IXS: as discussed above, the fact that the incident x-ray energy is several (>5, usually) orders of magnitude larger than the measured energy transfer leads to a complete decoupling between energy transfer and momentum transfer. This is in distinct contrast to INS where typically, especially if one wants good energy resolution, the neutron energy is comparable or only slightly larger than the energy transfer measured. Practically, this means that kinematic conservation laws only allow INS scans over a limited range of energy transfers at small Q, which, especially for liquids with higher sound speeds, may not extend over the full range of the acoustic mode energy. While there has been effort to exceed this limitation for INS (e.g. BRISP [19] ) that work had low rates so that thick samples were needed and the multiple-scattering background subtracted was much larger than the signal [20] . IXS, meanwhile, is largely background free, excepting relatively small contributions from sample containers and stray scattering that can be measured and subtracted. The kinematic limits of INS might be part of the reason that initial IXS work investigated disordered materials.

IXS work on liquids in Japan was facilitated by a strong community interested in liquid structure. This very naturally extends to dynamics as, for liquids, one can not separate the two: constant ps-scale motion, including diffusion, determines the structure. Thus, when IXS became available in Japan, there was significant interest in measuring liquids.

We briefly discuss the experimental setup for liquids. In general the tricky part of liquid experiments is making a sealed sample chamber that does not create large backgrounds, especially at the very low Qs that are often interesting for liquid work. For temperatures below ~400K this is not too difficult, and several designs with single crystal diamond or sapphire windows have been used. However as temperature increases, this becomes more difficult. Often we have used Hiroshima type sample cells which are precisely milled from single-crystal sapphire [21] as these allow temperatures in excess of 2000K. To achieve high temperatures one of several sample chambers have been used, including the Marburg chamber for modest pressures and (now) T<1400K, a local chamber [22] that can be used to nearly 2000K for samples in vacuum and also custom designs for higher pressure (up to 300 or even 1000 bar) and modestly high temperature, though those require special paperwork and licensing to conform to local regulations about high pressure gas environments.

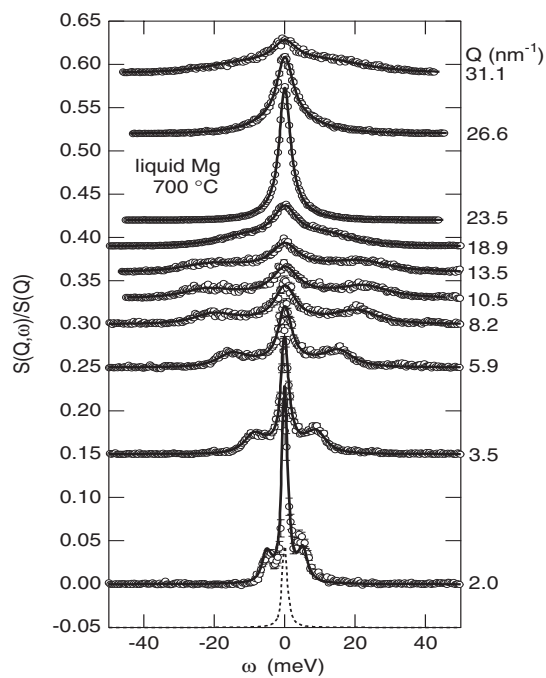

**Figure 2.** IXS spectra from liquid Mg as a function of energy transfer at different momentum transfers. See discussion in text. After [23] – the first published spectra from measurements at SPring-8.

Studies have included liquid metals such as magnesium [23] , silicon [24] , near critical mercury [25] gallium [26] , iron [27] , calcium and cadmium [28] and occasional molecular liquids including CCl$_4$ [29] near critical water [30] and methanol [31] , acetone [32] , water [33] , benzene [34] , *etc.* Studies of more complicated systems, including a quasi-crystal melt [35] , tellurium [36] , selenium [37] [38] , selenium-tellurium mixtures [39] , sulfur [40] , As$_2$Se$_3$ [41] , hydrated protein [42] , room-temperature ionic liquids [43] , have also been carried out. We discuss some results in more detail below, but also emphasize this is neither an exhaustive list of samples nor of papers about the samples mentioned.

To a first approximation, the spectra measured in IXS studies are largely similar across liquid systems (one example from [23] is shown in figure 2): at low Q (< 20 or 30 nm$^{-1}$) the dynamic structure factor has a strong quasi-elastic peak at zero energy transfer bracketed by the peaks of the stokes and anti-stokes contributions of the acoustic mode, while at higher Q the excitations generally merge into a broad peak centered near zero energy transfer. Whether the acoustic peaks are easily discernable or not at low Q depends on both the spectrometer resolution and the sample behavior, including the Landau-Placzek-ratio, the sound velocity, and the presence and strength of structural relaxations that, in effect, define the mesoscale liquid behavior. The example of figure 2 has relatively well defined acoustic peaks - other liquids, especially molecular liquids, often having acoustic modes that are only visible as weak shoulders on a logarithmic intensity scale. Usually the quasi-elastic and acoustic mode linewidth increases as Q increases (though more slowly than the $Q^2$ dependence of hydrodynamics). Then as one approaches the position of the first structure factor (SF) maximum the acoustic modes disperse inwards, to lower energy - the SF maximum acts like the second pseudo-brilloiun zone center in a solid. There the quasi-elastic mode also shows deGennes narrowing, becoming narrower at the SF maximum. These effects all evident by direct inspection of the example shown in figure 2.

Topics of interest in liquid dynamics include both the basic features and more sophisticated ones. Basic features include the direct acoustic mode dispersion, and the presence and intensity of relaxations. In general liquids show a phenomenon called "positive dispersion" or, sometimes, "fast sound", where the acoustic mode frequency at the mesoscopic Q values accessible by IXS (say, 1 to 8 nm$^{-1}$) increases more quickly with momentum transfer than expected from the macroscopic (ultrasonic) sound velocity. This may be interpreted as a visco-elastic affect where, as Q is increased and the sound frequency increases, the system response changes from viscous (or lossy as one expects in a liquid) to elastic (or more solid like). This can be considered as being a direct result of structural relaxation, and, in particular, Mountain [18] showed that adding a relaxation time in hydrodynamics both leads to a new quasi-elastic peak (relaxation) in the dynamical response and causes positive dispersion. Unfortunately, Mountain's work seems to be somewhat neglected in the world of IXS (and INS) work, possibly because the approximations needed to make the results tractable mean the simplified line shapes were not appropriate for careful analysis of spectra, but also probably because other methods of analysis, such as a generalized Langevin or memory function approach were more popular/trendy. Discussion of the memory function approach can be found in [4].

There are a variety of topics that have garnered interest in liquid dynamics and we mention a few briefly. This includes changes in acoustic dispersion associated with a metal-non-metal transition – as was measured specifically for near-critical mercury [25] but may be a general feature of liquid metal dynamics as, when T is increased, basically all liquid metals expand and become insulators. Another interesting feature in the critical region is the appearance of distinct liquid-like and gas-like contributions to the dynamics, with a cross-over at the Widom line, as observed for supercritical water [44] and in MD simulations of other systems [45]. There has been discussion about the possible appearance of transverse acoustic modes in spectra in several liquid metals [46–48] however, the evidence for that is indirect, as IXS probes only the longitudinal components of atomic motion. There is clear evidence of molecular or heavy particle modes from polymeric chain fragments in liquid selenium [49]. Investigation of the acoustic dispersion of several materials including liquid Bismuth [50], GeTe [51] and a phase change melt [52] have shown the acoustic dispersion deviates from a simple sinusoid, including a flat top region (see figure 3) as is postulated to be due a more complicated nearest neighbor bonding. It has been taken as evidence of a Peierls-like (short/long bond-length) distortion as is observed in some solids. There has been an effort to relate the detailed shape and molecular flexibility to the presence of fast sound in some liquids [53] The interpretation of liquid spectra can be complicated because the lines tend to be broad so one is sometimes very sensitive to the exact parametrization of the line shape. Thus, for example, the proper inclusion of the interaction term in fitting the spectra of liquid water allowed resolution of a long-standing controversy about the possible presence of an extra mode at low Q [17] : the extra mode was an artifact of fitting without the interaction (see figure 4).

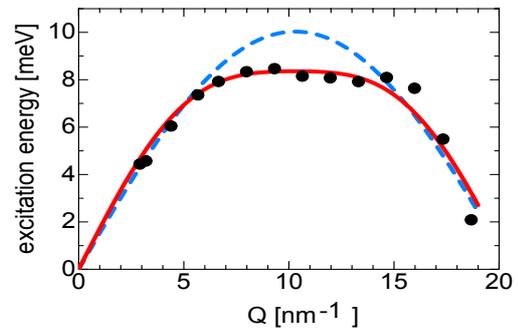

**Figure 3.** Measured dispersion of the acoustic mode of liquid Bi at 573K (points) as compared to a sin curve (dashed line) and a chain model with 2 force constants (solid line) simulating the response with a Peierls distortion. Based on results in [50]

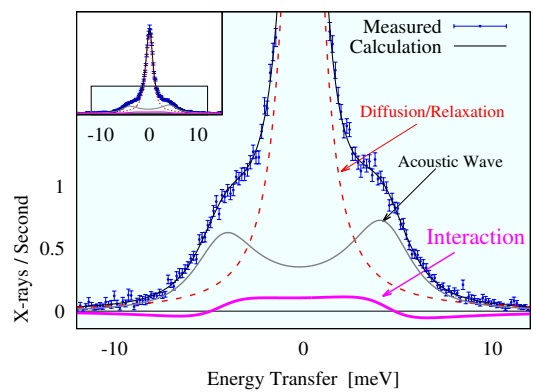

**Figure 4.** Spectrum from ambient water at Q=2.5 nm$^{-1}$ with ~0.9 meV resolution expanded to show the three different components, of eqn (5). The inset shows the full spectrum. Fits without the interaction added a weak mode at low energy that is not needed when the correct form is used. After [17].

Another approach to liquid dynamics has been to recover the Van Hove function by measuring over a large range of Q and E and transforming to real space and time. This has been done with IXS for water and some other molecular liquids [54–57] . It offers a different viewpoint on the dynamics, with, for example, an apparent mixing of the first and second neighbor motions in liquid water as seen in figure 5. An alternative method of processing also allows extraction of the self (or incoherent) part of the correlation function [58,59] as is difficult to do otherwise with x-ray scattering. We note a similar transformation has also been applied to INS data from liquid bismuth [60] .

Finally, in this section on disordered materials, we briefly touch on results for glasses. There have been fewer studies of glasses in Japan than liquids. Specific samples have included amorphous selenium [61] , $As_xSe_{1-x}$ $GeO_2$ [62] , silica [63] and some metal glasses [64–67] . Superficially the spectra of glasses appear similar to those of liquids, as seen in figure 6. However, there are two notable differences: the first is that the peak at zero energy is resolution limited (ie:

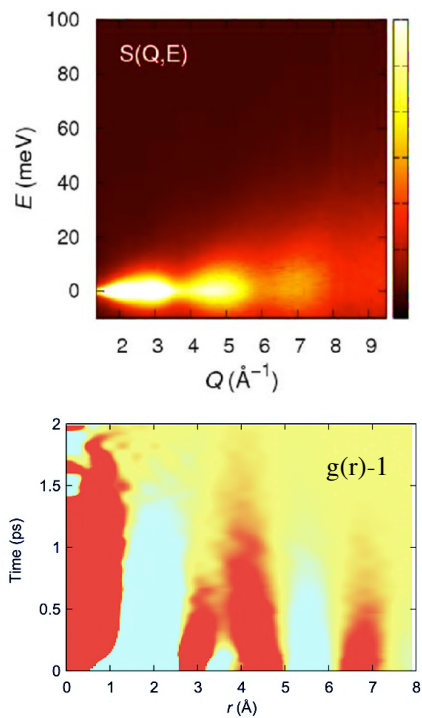

**Figure 5.** Dynamic structure factor and Van Hove correlation function for liquid water. After [54] .

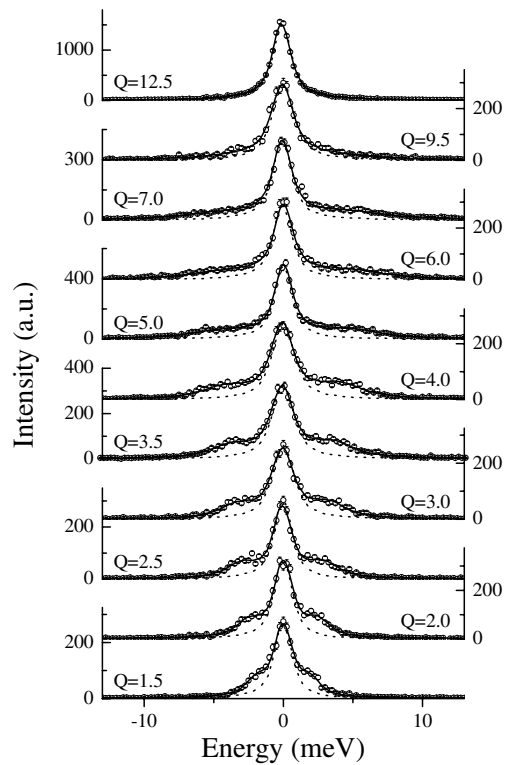

**Figure 6.** Spectra of amorphous selenium in ambient conditions. After [61]

elastic) on the meV scale, and the second is that in many cases a second mode appears clearly at low energies for, say, Q>4 nm$^{-1}$, as may be an indication of transverse dynamics. One notes that, being solids, glasses are expected to support transverse modes much more easily than liquids. This additional mode has been seen, e.g., in vitreous $GeO_2$ [62] and also in silica-based glasses [68,69] . Other areas of interest for glasses include the momentum dependence of the acoustic mode linewidth as it responds to disorder (e.g. [70]), the magnitude of the non-ergodicity parameter as might be related to fragility [71] , the possible presence of fast sound [64] and, of course, always, possible information about the boson peak.

## 6. Dynamics of Crystals

The meV-scale dynamics of crystalline materials as investigated by IXS has been extensively discussed in a review that updated in 2020 [1] . Therefore the discussion here will be very brief and more about instrumentation. IXS experiments on crystalline materials are easily possible between about 10 and 800K in a usual closed-cycle cryofurnace system. A Joule-Thomson stage system can also allow lower temperatures (~2K) to be reached if needed. Samples can be measured in a Bragg (reflection) geometry that will often give higher rates, but requires larger (mm scale) surfaces and limits the range of accessible momentum transfers to ones that are relatively close to the sample surface normal. Experiments can also be done in a Laue (transmission) geometry if the sample is thin enough (e.g. ~1 attenuation length thick). This usually gives access to a larger range of momentum transfers, but also reduces rates, typically by a factor of 2 or more relative to a Bragg geometry. Typically a comfortable size of a sample transverse to the x-ray beam is of ~1mm scale, but experiments have been done on ~0.3 to 0.5 mm scale samples. Below that size, one must consider carefully how to mount a sample, as most methods will put some other materials in the beam and give backgrounds. Unfortunately, possibly even because samples can be small, users too often do not investigate their sample ahead of experiments, and poor sample quality is the single most common cause of experimental failure. None-the-less, the majority of IXS experiments and publications are on crystalline systems.

## 7. Specific Geometries: DACs and Films

There are some specific types of experiments that are possible with IXS that are not possible using INS. These include measurements of samples in Diamond anvil cells (DACs) and measurements of thin films. As these are different than what is usually done in INS and as they both require somewhat specialized setups we discuss them separately. However, note that a broader survey of high pressure measurements can be found in [1].

Diamond anvil cells are used to generate pressures ~ 2 to >300 GPa, with lower pressures being tricky to control and higher pressures risking diamond breakage: the comfort region is, depending on cell design, something like 3 to 100 GPa. In order to get to higher pressures the culet diameter (the pressure side area) of the diamonds is reduced – whereas low pressures may be done with ~500 um culets, higher pressures are done with increasingly smaller culets, with, e.g., ~150 um culets up to ~ 150 GPa, and even smaller, to a few 10's of microns with special designs, at pressures > 150 GPa. As the sample size is usually 1/3 (or less) of the cullet size, and as the sample thickness also is reduced at higher pressure, DAC experiments quickly become more difficult as pressure is increased: pressures < 100 GPa seem to be possible in most cases, but above that they get more difficult quickly. There are also two laser heating setups at SPring-8 that are designed for samples in DACs, and these belong to specific user groups.

There are essentially 3 types of DAC experiments: investigation of acoustic mode dispersion in powder/polycrystalline or liquid samples at low Q, investigation of acoustic mode dispersion in crystalline samples at higher Q giving the full elasticity tensor, and investigation of optical modes in single crystal samples at high Q. Most work at SPring-8 focusses on the first two experiments for geoscience: sound velocities and elasticity in known conditions is critical information needed to interpret seismological measurements of the earth's interior.

A variety of specialized instrumentation has been developed to do this work. Chief among these is a multilayer KB focusing setup at BL43LXU which provides a 5 micron diameter x-ray beam size [72,73], the smallest now available at any IXS spectrometer. In addition, a "Soller screen" is used to help reduce backgrounds [72,73]. Placing a few micron thick sample between two ~1.5 mm thick diamonds leads to a nasty S/N problem: the Soller screen reduces the background from the diamonds. Meanwhile the measurement of elasticity from single crystals mostly proceeds according to [74].

After years of R&D, recent success stories include the measurement of room temperature iron to ~300 GPa [75], the determination of a new primary pressure scale to > 200 GPa [76], and the determination of an equation of state for liquid iron [77]. Meanwhile single crystal measurements of elasticity have been extended to > 50 GPa [78].

Thin film measurements are interesting because there are a variety of materials that are essentially only available in thin-film form – including "bulk" materials that are stabilized by the substrate or materials where one is creating some sort of periodic or semi-periodic layered structure as may be especially interesting for thermal engineering. However, the obvious problem is that the samples are thin, with an exceptionally thick film being ~ 1 micron thick, and more typical thicknesses being ~100 nm or less. Then, for IXS, one typically employs a grazing incidence geometry, where the beam is incident nearly parallel to the surface as this helps to increase the path in the sample and hence the signal. However, this requires both rather careful calculations that properly take account of the tilting of the incident x-ray beam by upstream focusing elements (e.g. [79]), and additional stages and specialized setups to deal with proper alignment at grazing incidence. This work has largely been done at BL35XU. The first published thin film work was done to investigate the phonon structure InGaAs/GaAsP quantum well structures [80], with later work investigating "bulk" ScN [81] and a HfN/HfSc superlattice [82] Work is continuing in several directions.

## 8. Future Plans

The IXS spectrometers at SPring-8 are mature and robust instruments and indeed now offer the best available performance, world-wide. Nevertheless, there is ongoing work to improve their performance. It has been possible to improve the resolution in the workhorse setup at BL43LXU to between 1.2 and 1.3 meV on most analyzers using carefully employed area detectors [83,84] – a similar improvement is planned for BL35XU. Meanwhile work continues to improve sample environments, such as extending the range of temperatures possible for thin film measurements. Finally, recent work showed it was possible, in principle to attain <0.4 meV resolution [85] , and, while the setup is severely flux limited, it is being improved to make sample measurements more feasible.

The planned upgrade from SPring-8 to SPring-8-II will make some changes in the IXS capabilities, but, while the upgrade is very important for other experiments, the impact will be relatively small for IXS. This is because present IXS experiments are more flux limited than brilliance limited, and the SPring-8-II upgrade primarily improves brilliance. Thus, for IXS, one mainly expects that the focal spot sizes with the SPring-8-II source will be reduced in the horizontal and better momentum transfer, Q, resolution (<0.1 nm$^{-2}$) may become easier. We note that the present IDs used for IXS are highly optimized for operation near 22 keV (the energy of the workhorse ~1 meV resolution setups) the change in IDs needed for SPring-8-II should result in higher flux at higher energy (25-30 keV) as may significantly improve count-rates at higher (sub-meV and even sub-0.4 meV) resolution.

**Alfred Q.R. Baron** 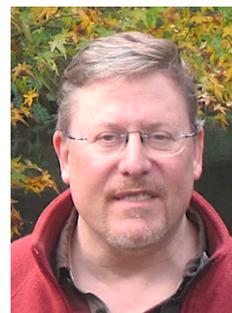
Alfred Baron received a Ph.D. doing nuclear resonant scattering of synchrotron radiation (NRS) and synchrotron instrumentation in the Department of Applied Physics at Stanford University. He then moved to ESRF to work at the first 3$^{rd}$ generation NRS beamline where he contributed to many pioneering experiments, including investigations of x-ray coherence in NRS and the development of a new approach using NRS to probe atomic dynamics of non-resonant samples on the neV scale - time domain interferometry – as is roughly comparable to neutron spin echo work. He moved to SPring-8 in 1997 where he built Japan's first meV IXS spectrometer. He shifted from JASRI to RIKEN in 2006, creating the Materials Dynamics Laboratory, and building the world's most powerful beamline for meV IXS, the RIKEN Quantum NanoDynamics Beamline, BL43LXU. His main appointment is as the Director of the Materials Dynamics Group of the RIKEN SPring-8 Center, and he also serves as the Director of the Precision Spectroscopy Division in the Center for Synchrotron Radiation Research (CSRR) of JASRI. His scientific interests continue to include many aspects of atomic dynamics and development of world-leading instrumentation. In recent years his interest has shifted to focus more on the dynamics of disordered materials as a field with opportunity for significant progress, especially taking advantage of the spectrometers at SPring-8

Affiliation：RIKEN SPring-8 Center
e-mail：baron@spring8.or.jp
*Research Fields：Atomic Dynamics, Inelastic x-ray scattering, Nuclear Resonant Scattering, Synchrotron Radiation Instrumentation, X-ray Optics, X-ray Detectors*